\documentclass{article}
\usepackage{amsmath}

\providecommand{\keywords}[1]{\text{\textbf{Keywords: }} #1}

\begin{document}

\title{Neutrino-assisted fermion-boson transitions }
\author{Andrzej Okni\'{n}ski \\
%EndAName
Physics Division, Politechnika \'{S}wi\c{e}tokrzyska, \\
Al. 1000-lecia PP 7, 25-314 Kielce, Poland}
\maketitle

\begin{abstract}
We study fermion-boson transitions. Our approach is based on the $3\times 3$
subequations of Dirac and Duffin-Kemmer-Petiau equations, which link these
equations. We demonstrate that free Dirac equation can be invertibly
converted to spin-$0$ Duffin-Kemmer-Petiau equation in presence of a
neutrino field. We also show that in special external fields, upon assuming
again existence of a neutrino (Weyl) spinor, the Dirac equation can be
transformed reversibly to spin-$0$ Duffin-Kemmer-Petiau equation. We argue
that such boson-fermions transitions are consistent with the main channel of
pion decay.
\end{abstract}


\section{Introduction}

\label{introduction}

There are many ideas connected with fermion-boson (FB) analogies in the
literature. For example, there is FB equivalence, FB duality, FB
transmutations, to name a few. There are also intermediate statistics -
parastatistics and anyons. There is, finally, supersymmetry. More on these
ideas can be found in Refs. \cite%
{Garbaczewski1985,Wilczek1990,Stone1994,Simulik2005}, see also \cite%
{Garbaczewski1975,Jackiw1991,Plyushchay1991,Horvathy2010} and \cite%
{Okninski2014}. It seems, however, that a broader and unifying picture is
still missing.

Important step in understanding FB analogy was made by Polyakov who
discovered possibility of fermion-boson transmutation of elementary
excitations of a scalar field interacting with the topological Chern-Simons
term in ($2+1$) dimensions \cite{Polyakov1988}. Recently, the smooth and
controlled evolution from a fermionic Bardeen-Cooper-Schrieffer (BCS)
superfluid state to a molecular Bose-Einstein condensate (BEC)\ has been
realized in ultracold Fermi gases \cite{Zwerger2012}. On the other hand, we
have shown recently that solutions of the Dirac equation can be transformed
in the non-interacting case, assuming existence of a constant spinor, to
solutions of the spin-$0$ Duffin-Kemmer-Petiau (DKP) equation and vice versa 
\cite{Okninski2014}. Possible analogy between BCS-BEC transition and our
findings is motivation of our work.

In the present paper we generalize results of \cite{Okninski2014} in two
directions. Firstly, we generalize the fermion-boson transformation
connecting solutions of free Dirac and spin-$0$ DKP equations in presence of
Weyl spinor. Secondly, we construct analogous transformation in presence of
external fields.

The paper is organized as follows. In the next Section we review $3\times 3$
subequations of the Dirac and the spin-$0$ DKP equations in special external
fields and we raise the problem of their covariance. Covariance of these
equations was established in \cite{Okninski2012} while the problem of
covariance of their solutions was solved, to some extent, in Ref. \cite%
{Okninski2014}.

New results are described in Sections \ref{transition}, \ref{mechanism}. In
Section \ref{transition} we start with the free Dirac equation and, assuming
existence of a Weyl spinor, we derive the spin-$0$ DKP equation, improving
our construction described in \cite{Okninski2014}. It is suggested that the
mechanism of boson to massive fermion and massless neutrino transition is
related to pion decay. In Section \ref{mechanism} we show that in external
longitudinal fields, upon assuming again existence of a Weyl spinor, the
Dirac equation can be transformed to a set of two $3\times 3$ equations in
longitudinal fields, similar, but not identical, to equations derived in 
\cite{Okninski2012}. Finally, we show that if we switch over to crossed
fields we arrive at $3\times 3$ equations which are equivalent to the spin-$%
0 $ DKP equations. We discuss our results in the last Section with special
emphasis on pion decay. In what follows we use notation and conventions
described in \cite{Okninski2011,Okninski2012}.

\section{Subequations of Dirac and DKP equations in external fields}

\label{subequations}

The Dirac equation in external field can be written in spinor notation as 
\cite{Berestetskii1971}:

\begin{equation}
\left. 
\begin{array}{r}
\pi ^{A\dot{B}}\eta _{\dot{B}}=m\xi ^{A}\smallskip \\ 
\pi _{A\dot{B}}\xi ^{A}=m\eta _{\dot{B}}%
\end{array}%
\right\} ,  \label{Dirac1}
\end{equation}%
where $\pi ^{A\dot{B}}$ is defined as $\pi ^{A\dot{B}}=\left( \sigma ^{0}\pi
^{0}+\overrightarrow{\sigma }\cdot \overrightarrow{\pi }\right) ^{A\dot{B}}$%
, $\pi ^{\mu }=p^{\mu }-qA^{\mu }$, and $\pi _{1\dot{1}}=\pi ^{2\dot{2}}$, $%
\pi _{1\dot{2}}=-\pi ^{2\dot{1}}$, $\pi _{2\dot{1}}=-\pi ^{1\dot{2}}$, $\pi
_{2\dot{2}}=\pi ^{1\dot{1}}$. Eq. (\ref{Dirac1}) can be written in the
spinor representation of $\gamma $ matrices as $\gamma ^{\mu }\pi _{\mu
}\Psi =m\Psi $, where $\Psi =\left( \xi ^{1},\xi ^{2},\eta _{\dot{1}},\eta _{%
\dot{2}}\right) ^{T}$.

In longitudinal fields \cite{Bagrov1990} Eq. (\ref{Dirac1}) can be splitted
into two $3\times 3$ subequations \cite{Okninski2012}:%
\begin{equation}
\left. 
\begin{array}{rcl}
\pi ^{1\dot{1}}\eta _{\dot{1}} & = & m\psi _{\dot{1}}^{1\dot{1}} \\ 
\pi ^{2\dot{1}}\eta _{\dot{1}} & = & m\psi _{\dot{1}}^{2\dot{1}} \\ 
\pi ^{2\dot{2}}\psi _{\dot{1}}^{1\dot{1}}-\pi ^{1\dot{2}}\psi _{\dot{1}}^{2%
\dot{1}} & = & m\eta _{\dot{1}}%
\end{array}%
\right\} ,  \label{const1b}
\end{equation}%
\begin{equation}
\left. 
\begin{array}{rcl}
\pi ^{1\dot{2}}\eta _{\dot{2}} & = & m\psi _{\dot{2}}^{1\dot{2}} \\ 
\pi ^{2\dot{2}}\eta _{\dot{2}} & = & m\psi _{\dot{2}}^{2\dot{2}} \\ 
-\pi ^{2\dot{1}}\psi _{\dot{2}}^{1\dot{2}}+\pi ^{1\dot{1}}\psi _{\dot{2}}^{2%
\dot{2}} & = & m\eta _{\dot{2}}%
\end{array}%
\right\} .  \label{const2b}
\end{equation}%
Each of these equations can be written in covariant form \cite%
{Okninski2011,Okninski2012} yet some components of spinor $\psi _{\dot{C}}^{A%
\dot{B}}$ are missing. This problem, mentioned in the Introduction, will be
solved in the next Section.

The DKP equation in the interacting case can be written within spinor
formalism as:

\begin{equation}
\left. 
\begin{array}{ccc}
\pi ^{A\dot{B}}\psi & = & m\psi ^{A\dot{B}} \\ 
\pi _{A\dot{B}}\psi ^{A\dot{B}} & = & 2m\psi%
\end{array}%
\right\} .  \label{KDP-s0-2}
\end{equation}

In crossed fields \cite{Bagrov1990} we can write Eqs.(\ref{KDP-s0-2}) as a
set of two equations \cite{Okninski2012}: 
\begin{equation}
\left. 
\begin{array}{r}
\pi ^{1\dot{1}}\psi =m\psi ^{1\dot{1}} \\ 
\pi ^{2\dot{1}}\psi =m\psi ^{2\dot{1}} \\ 
\pi _{1\dot{1}}\psi ^{1\dot{1}}+\pi _{2\dot{1}}\psi ^{2\dot{1}}=m\psi%
\end{array}%
\right\} ,  \label{const-s0-1}
\end{equation}%
\begin{equation}
\left. 
\begin{array}{r}
\pi ^{1\dot{2}}\psi =m\psi ^{1\dot{2}} \\ 
\pi ^{2\dot{2}}\psi =m\psi ^{2\dot{2}} \\ 
\pi _{1\dot{2}}\psi ^{1\dot{2}}+\pi _{2\dot{2}}\psi ^{2\dot{2}}=m\psi%
\end{array}%
\right\} ,  \label{const-s0-2}
\end{equation}%
each of which describes particle with mass $m$. Eq. (\ref{KDP-s0-2}) and the
set of two equations (\ref{const-s0-1}), (\ref{const-s0-2}) are equivalent.

The $3\times 3$ equations (\ref{const-s0-1}), (\ref{const-s0-2}) and (\ref%
{const1b}), (\ref{const2b}) are similar. However, they differ in spinor
contents and Eqs. (\ref{const1b}), (\ref{const2b}) involve longitudinal
fields while Eqs. (\ref{const-s0-1}), (\ref{const-s0-2}) correspond to
crossed fields.

\section{Generalized fermion-boson transition in non-interacting case}

\label{transition}

The problem of missing components of spinor $\psi _{\dot{C}}^{A\dot{B}}$,
mentioned in the previous Section, is rather serious because it means that
theory is not fully covariant. The problem was solved in Ref. \cite%
{Okninski2014} assuming that $\eta _{\dot{B}}\left( x\right) =\hat{\alpha}_{%
\dot{B}}\,\chi \left( x\right) $, where $\hat{\alpha}_{\dot{B}}$ was a
constant spinor. This ansatz for $\eta _{\dot{B}}\left( x\right) $ leads,
however, to difficulties of another kind since constant spinors do not
appear in nature (although constant Grassman spinors are postulated in some
variants of supersymmetrical theories). To solve both problems at once we
make a more general assumption:%
\begin{equation}
\eta _{\dot{B}}\left( x\right) =\alpha _{\dot{B}}\left( x\right) \chi \left(
x\right) ,  \label{neutrino1F}
\end{equation}%
where $\alpha _{\dot{B}}\left( x\right) $ is a two-component neutrino
spinor, i.e. it fulfills the Weyl equation, $p^{A\dot{B}}\alpha _{\dot{B}%
}\left( x\right) =0$. We note that $p^{A\dot{B}}\alpha _{\dot{B}}\chi =\chi
\,p^{A\dot{B}}\alpha _{\dot{B}}+\alpha _{\dot{B}}\,p^{A\dot{B}}\chi =\alpha
_{\dot{B}}\,p^{A\dot{B}}\chi $ and thus we can rewrite the free Dirac
equation as:

\begin{equation}
\left. 
\begin{array}{r}
\alpha _{\dot{B}}p^{A\dot{B}}\chi =m\xi ^{A}\smallskip \\ 
p_{A\dot{B}}\xi ^{A}=m\alpha _{\dot{B}}\chi%
\end{array}%
\right\} .  \label{Dirac2F}
\end{equation}

Equation (\ref{Dirac2F}) can be further written as:

\begin{equation}
\left. 
\begin{array}{r}
\alpha _{\dot{1}}p^{1\dot{1}}\chi =m\psi _{\dot{1}}^{1\dot{1}} \\ 
\alpha _{\dot{2}}p^{1\dot{2}}\chi =m\psi _{\dot{2}}^{1\dot{2}} \\ 
\alpha _{\dot{1}}p^{2\dot{1}}\chi =m\psi _{\dot{1}}^{2\dot{1}} \\ 
\alpha _{\dot{2}}p^{2\dot{2}}\chi =m\psi _{\dot{2}}^{2\dot{2}} \\ 
\left( p_{1\dot{1}}\psi _{\dot{1}}^{1\dot{1}}+p_{2\dot{1}}\psi _{\dot{1}}^{2%
\dot{1}}\right) +\left( p_{1\dot{1}}\psi _{\dot{2}}^{1\dot{2}}+p_{2\dot{1}%
}\psi _{\dot{2}}^{2\dot{2}}\right) =m\alpha _{\dot{1}}\chi \smallskip \\ 
\left( p_{1\dot{2}}\psi _{\dot{1}}^{1\dot{1}}+p_{2\dot{2}}\psi _{\dot{1}}^{2%
\dot{1}}\right) +\left( p_{1\dot{2}}\psi _{\dot{2}}^{1\dot{2}}+p_{2\dot{2}%
}\psi _{\dot{2}}^{2\dot{2}}\right) =m\alpha _{\dot{2}}\chi%
\end{array}%
\right\} ,  \label{Dirac3F}
\end{equation}%
where%
\begin{equation}
\left. 
\begin{array}{c}
\psi _{\dot{1}}^{1\dot{1}}+\psi _{\dot{2}}^{1\dot{2}}=\xi ^{1} \\ 
\psi _{\dot{1}}^{2\dot{1}}+\psi _{\dot{2}}^{2\dot{2}}=\xi ^{2}%
\end{array}%
\right\} .  \label{conditions1F}
\end{equation}

To reduce number of spinor components we demand that: 
\begin{equation}
\psi _{\dot{B}}^{C\dot{D}}=\alpha _{\dot{B}}\left( x\right) \chi ^{C\dot{D}%
}\left( x\right) ,  \label{neutrino2F}
\end{equation}%
with the same spinor $\alpha _{\dot{B}}\left( x\right) $, fulfilling the
Weyl equation $p^{A\dot{B}}\alpha _{\dot{B}}\left( x\right) =0$.
Substituting (\ref{neutrino2F}) into Eq. (\ref{Dirac3F}) we obtain: 
\begin{equation}
\left. 
\begin{array}{r}
\alpha _{\dot{1}}p^{1\dot{1}}\chi =m\alpha _{\dot{1}}\chi ^{1\dot{1}} \\ 
\alpha _{\dot{1}}p^{2\dot{1}}\chi =m\alpha _{\dot{1}}\chi ^{2\dot{1}} \\ 
\left( p_{1\dot{1}}\alpha _{\dot{1}}\chi ^{1\dot{1}}+p_{2\dot{1}}\alpha _{%
\dot{1}}\chi ^{2\dot{1}}\right) +\left[ p_{1\dot{1}}\alpha _{\dot{2}}\chi ^{1%
\dot{2}}+p_{2\dot{1}}\alpha _{\dot{2}}\chi ^{2\dot{2}}\right] =m\alpha _{%
\dot{1}}\chi \\ 
\alpha _{\dot{2}}p^{1\dot{2}}\chi =m\alpha _{\dot{2}}\chi ^{1\dot{2}} \\ 
\alpha _{\dot{2}}p^{2\dot{2}}\chi =m\alpha _{\dot{2}}\chi ^{2\dot{2}} \\ 
\left[ p_{1\dot{2}}\alpha _{\dot{1}}\chi ^{1\dot{1}}+p_{2\dot{2}}\alpha _{%
\dot{1}}\chi ^{2\dot{1}}\right] +\left( p_{1\dot{2}}\alpha _{\dot{2}}\chi ^{1%
\dot{2}}+p_{2\dot{2}}\alpha _{\dot{2}}\chi ^{2\dot{2}}\right) =m\alpha _{%
\dot{2}}\chi%
\end{array}%
\right\} .  \label{Dirac4F}
\end{equation}%
Since the Weyl spinor $\alpha _{\dot{B}}\left( x\right) $ is arbitrary,
equations $p^{A\dot{B}}\chi=m\chi ^{A\dot{B}}$, defining components of $\chi
^{A\dot{B}}$, follow immediately. We can thus remove spinor components $%
\alpha _{\dot{1}}$, $\alpha _{\dot{2}}$ from equations defining $\chi ^{A%
\dot{B}}$.

We have shown that for constant spinor, $\alpha _{\dot{B}}\left( x\right) =%
\hat{\alpha}_{\dot{B}}$, the system of equations (\ref{Dirac4F}) splits into
two $3\times 3$ equations \cite{Okninski2014}. We are going to find
conditions enabling similar splitting for the Weyl spinor $\alpha _{\dot{B}%
}\left( x\right) $.

We write the first and the second term in square brackets, $\left[ 1\right] $
and $\left[ 2\right] $, respectively, in form: 
\begin{subequations}
\label{SQF}
\begin{align}
\left[ 1\right] & =\chi ^{1\dot{2}}p_{1\dot{1}}\alpha _{\dot{2}}+\chi ^{2%
\dot{2}}p_{2\dot{1}}\alpha _{\dot{2}}+\tfrac{1}{m}\alpha _{\dot{2}}\left(
p_{1\dot{1}}p^{1\dot{2}}+p_{2\dot{1}}p^{2\dot{2}}\right) \chi ,  \label{sq1F}
\\
\left[ 2\right] & =\chi ^{1\dot{1}}p_{1\dot{2}}\alpha _{\dot{1}}+\chi ^{2%
\dot{1}}p_{2\dot{2}}\alpha _{\dot{1}}+\tfrac{1}{m}\alpha _{\dot{1}}\left(
p_{1\dot{2}}p^{1\dot{1}}+p_{2\dot{2}}p^{2\dot{1}}\right) \chi ,  \label{sq2F}
\end{align}%
where equations $p^{A\dot{B}}\chi =m\chi ^{A\dot{B}}$ have been used.

Since both terms in (\ref{SQF}), proportional to $\frac{1}{m}$, vanish
identically we can decouple equations (\ref{Dirac4F}) obtaining: 
\end{subequations}
\begin{eqnarray}
&&\left. 
\begin{array}{r}
p^{1\dot{1}}\chi =m\chi ^{1\dot{1}} \\ 
p^{2\dot{1}}\chi =m\chi ^{2\dot{1}} \\ 
p_{1\dot{1}}\chi ^{1\dot{1}}+p_{2\dot{1}}\chi ^{2\dot{1}}=m\chi%
\end{array}%
\right\} ,\medskip  \label{decoupled1F} \\
&&\left. 
\begin{array}{r}
p^{1\dot{2}}\chi =m\chi ^{1\dot{2}} \\ 
p^{2\dot{2}}\chi =m\chi ^{2\dot{2}} \\ 
p_{1\dot{2}}\chi ^{1\dot{2}}+p_{2\dot{2}}\chi ^{2\dot{2}}=m\chi%
\end{array}%
\right\} ,  \label{decoupled2F}
\end{eqnarray}%
where unnecessary components $\alpha _{\dot{1}}$, $\alpha _{\dot{2}}$ have
been removed, provided that the following equations are fulfilled:

\begin{equation}
\left. 
\begin{array}{r}
\left( \chi ^{1\dot{1}}p_{1\dot{1}}\alpha _{\dot{1}}+\chi ^{2\dot{1}}p_{2%
\dot{1}}\alpha _{\dot{1}}\right) +\left( \chi ^{1\dot{2}}p_{1\dot{1}}\alpha
_{\dot{2}}+\chi ^{2\dot{2}}p_{2\dot{1}}\alpha _{\dot{2}}\right) =0 \\ 
\left( \chi ^{1\dot{1}}p_{1\dot{2}}\alpha _{\dot{1}}+\chi ^{2\dot{1}}p_{2%
\dot{2}}\alpha _{\dot{1}}\right) +\left( \chi ^{1\dot{2}}p_{1\dot{2}}\alpha
_{\dot{2}}+\chi ^{2\dot{2}}p_{2\dot{2}}\alpha _{\dot{2}}\right) =0%
\end{array}%
\right\} .  \label{conditions3F}
\end{equation}

Taking into account the form of solutions of the Weyl equation, $\alpha _{%
\dot{B}}\left( x\right) =\hat{\alpha}_{\dot{B}}e^{ik\cdot x}$, where $\hat{%
\alpha}_{\dot{B}}$ is a constant spinor and $k^{\mu }k_{\mu }=0$,\ we
rewrite Eqs. (\ref{conditions3F}) in form:

\begin{equation}
\left. 
\begin{array}{r}
k_{1\dot{1}}\varphi ^{1}+k_{2\dot{1}}\varphi ^{2}=0 \\ 
k_{1\dot{2}}\varphi ^{1}+k_{2\dot{2}}\varphi ^{2}=0%
\end{array}%
\right\} ,  \label{conditionsF}
\end{equation}%
where%
\begin{equation}
\left. 
\begin{array}{c}
\varphi ^{1}=\hat{\alpha}_{\dot{1}}\chi ^{1\dot{1}}+\hat{\alpha}_{\dot{2}%
}\chi ^{1\dot{2}} \\ 
\varphi ^{2}=\hat{\alpha}_{\dot{1}}\chi ^{2\dot{1}}+\hat{\alpha}_{\dot{2}%
}\chi ^{2\dot{2}}%
\end{array}%
\right\} .  \label{definitionsF}
\end{equation}

Non-zero solutions $\varphi ^{1}$, $\varphi ^{2}$ are possible if
determinant of the system of equations (\ref{conditionsF}) is zero. The
determinant, $k_{1\dot{1}}k_{2\dot{2}}-k_{1\dot{2}}k_{2\dot{1}}=k^{\mu
}k_{\mu }$, vanishes in two physically distinct cases: for $k^{\mu }=0$ and
for $k^{\mu }k_{\mu }=0$. In the first case the spinor $\alpha _{\dot{B}%
}\left( x\right) $ is constant, $\alpha _{\dot{B}}\left( x\right) =\hat{%
\alpha}_{\dot{B}}$, and no restrictions are imposed on $\varphi ^{1}$, $%
\varphi ^{2}$. The fermion-boson transformation in presence of a constant
spinor was investigated in the non-interacting case in \cite{Okninski2014}.

We consider now the second possibility. Since we have assumed that $\alpha _{%
\dot{B}}\left( x\right) =\hat{\alpha}_{\dot{B}}e^{ik\cdot x}$ is a solution
of the Weyl equation, the condition $k^{\mu }k_{\mu }=0$ is fulfilled.
Moreover, equations (\ref{decoupled1F}), (\ref{decoupled2F}) are the set of
two $3\times 3$ equations equivalent to the spin-$0$ DKP equation. Therefore 
$\chi $ fulfills the Klein-Gordon equation. Thus $\chi \left( x\right)
=Ce^{il\cdot x}$ where $l^{\mu }$ is a four-vector, $l^{\mu }l_{\mu }=m^{2}$.

Since the Weyl equation as well as the set of equations (\ref{decoupled1F}),
(\ref{decoupled2F}) are covariant we can consider special reference frames.
In a frame $k^{\mu }=\left( 1,0,0,-1\right) $, $l^{\mu }=\left(
m,0,0,0\right) $, we have $\hat{\alpha}_{\dot{1}}=0$, $\chi ^{1\dot{2}} = 0$%
, $\varphi ^{1}=0$, $k_{2\dot{2}}=0$ and it follows that equations (\ref%
{conditionsF}), (\ref{definitionsF}) are fulfilled.

We have thus described invertible transition from the free Dirac equation
for a spin-$\frac{1}{2}$ fermion, in presence of a massless spin-$\frac{1}{2}
$ fermion, described by a dotted Weyl spinor, to the free DKP equation for a
spin-$0$ boson. Indeed, starting from Eqs. (\ref{decoupled1F}), (\ref%
{decoupled2F}) and assuming conditions (\ref{conditions3F}), which as has
been stated above can be easily fulfilled, we can return to equation (\ref%
{Dirac2F}) and, finally, to the Dirac equation (\ref{Dirac1}) in
non-interacting case. Hence the inverse transformation, from boson to
fermions, can be written as 
\begin{equation}
B\longrightarrow \left( F\,\bar{\nu}_{F}\right)  \label{decayF1}
\end{equation}%
where B an F stand for boson and fermion, respectively, while $\bar{\nu}_{F}$
denotes antineutrino associated with the fermion F. In this reaction $\left(
F\, \bar{\nu}_{F}\right)$ is a two-fermion composite state. The process (\ref%
{decayF1}) seems to correspond to the first stage of the main channel of
pion decay \cite{Beringer2012}: 
\begin{equation}
\pi ^{-}\longrightarrow \left( \mu ^{-}\,\bar{\nu}_{\mu }\right)
\longrightarrow \mu ^{-}+\bar{\nu}_{\mu }  \label{decayF2}
\end{equation}
where we have postulated formation of the intermediate complex $\left( \mu
^{-}\,\bar{\nu}_{\mu}\right)$.

Indeed, the boson to fermions transformation, constructed in this Section,
cannot describe direct decay of a pion into muon and neutrino. This follows
from the fact that masses of a boson particle and a fermion, into which it
transforms in presence of a neutrino field, are equal in our theory.
Therefore, the present formalism seems to apply to the first stage of the
reaction (\ref{decayF2}) only, with postulated formation of the intermediate
complex state $\left( \mu ^{-}\,\bar{\nu}_{\mu }\right) $ of mass $m=m_{\pi
^{-}}=139.570$ MeV$/c^{2}$. In the second stage of the process (\ref{decayF2}%
) the complex state decays into muon of mass $m_{\mu ^{-}}=105.658$ MeV$%
/c^{2}$ and massless neutrino, the energy excess $\left( m_{\pi ^{-}}-m_{\mu
^{-}}\right) c^{2}$ converted into neutrino (mainly) and muon kinetic
energies. We shall comment on the suggested reaction in the last Section.

\section{Fermion-boson transition induced by change-over of external fields}

\label{mechanism}

In this Section we carry out splitting of the Dirac equation in the
interacting case. Let us assume as before the ansatz (\ref{neutrino1F}) and $%
p^{A\dot{B}}\alpha _{\dot{B}}\left( x\right) =0$. Computing $\pi ^{A\dot{B}%
}\eta _{\dot{B}}$ we get:%
\begin{equation}
\left( p-qA\right) ^{A\dot{B}}\alpha _{\dot{B}}\chi =\left( p^{A\dot{B}
}\alpha _{\dot{B}}\right) \chi +\alpha _{\dot{B}}\left( p^{A\dot{B}}\chi
\right) -qA^{A\dot{B}}\alpha _{\dot{B}}\chi =\alpha _{\dot{B}}\pi ^{A\dot{B}
}\chi ,  \label{computation}
\end{equation}%
and we can rewrite the Dirac equation (\ref{Dirac1}) as:%
\begin{equation}
\left. 
\begin{array}{r}
\alpha _{\dot{B}}\pi ^{A\dot{B}}\chi =m\xi ^{A}\smallskip \\ 
\pi _{A\dot{B}}\xi ^{A}=m\alpha _{\dot{B}}\chi%
\end{array}%
\right\} ,  \label{Dirac2}
\end{equation}%
where we assume that external field is longitudinal, i.e. fulfills
conditions $\left[ \pi ^{0}\pm \pi ^{3},\pi ^{1}\pm i\pi ^{2}\right] =0$.

We can now repeat all steps described in the preceding Section arriving at
equivalent of Eq. (\ref{SQF}): 
\begin{subequations}
\label{SQ}
\begin{align}
\left[ 1\right] & =\chi ^{1\dot{2}}p_{1\dot{1}}\alpha _{\dot{2}}+\chi ^{2%
\dot{2}}p_{2\dot{1}}\alpha _{\dot{2}}+\tfrac{1}{m}\alpha _{\dot{2}}\left(
\pi _{1\dot{1}}\pi ^{1\dot{2}}+\pi _{2\dot{1}}\pi ^{2\dot{2}}\right) \chi ,
\label{sq1} \\
\left[ 2\right] & =\chi ^{1\dot{1}}p_{1\dot{2}}\alpha _{\dot{1}}+\chi ^{2%
\dot{1}}p_{2\dot{2}}\alpha _{\dot{1}}+\tfrac{1}{m}\alpha _{\dot{1}}\left(
\pi _{1\dot{2}}\pi ^{1\dot{1}}+\pi _{2\dot{2}}\pi ^{2\dot{1}}\right) \chi .
\label{sq2}
\end{align}

We note now that terms in rounded brackets vanish identically in
longitudinal fields and we get: 
\end{subequations}
\begin{eqnarray}
&&\left. 
\begin{array}{r}
\pi ^{1\dot{1}}\chi =m\chi ^{1\dot{1}} \\ 
\pi ^{2\dot{1}}\chi =m\chi ^{2\dot{1}} \\ 
\pi _{1\dot{1}}\chi ^{1\dot{1}}+\pi _{2\dot{1}}\chi ^{2\dot{1}}=m\chi%
\end{array}%
\right\} ,\medskip  \label{decoupled1} \\
&&\left. 
\begin{array}{r}
\pi ^{1\dot{2}}\chi =m\chi ^{1\dot{2}} \\ 
\pi ^{2\dot{2}}\chi =m\chi ^{2\dot{2}} \\ 
\pi _{1\dot{2}}\chi ^{1\dot{2}}+\pi _{2\dot{2}}\chi ^{2\dot{2}}=m\chi%
\end{array}%
\right\} ,  \label{decoupled2}
\end{eqnarray}%
provided that again conditions (\ref{conditions3F}) are fulfilled but now $%
\chi ^{A\dot{B}},\,\chi $ are solutions of Eqs. (\ref{decoupled1}), (\ref%
{decoupled2}) in longitudinal fields.

Since solutions of the Weyl equation are of form $\alpha _{\dot{B}}\left(
x\right) =\hat{\alpha}_{\dot{B}}e^{ik\cdot x}$, where $\hat{\alpha}_{\dot{B}%
} $ is a constant spinor and $k^{\mu }k_{\mu }=0$,\ we rewrite Eqs. (\ref%
{conditions3F}) as equations (\ref{conditionsF}), (\ref{definitionsF}). It
follows that determinant of Eqs. (\ref{conditionsF}), $k_{1\dot{1}}k_{2\dot{2%
}}-k_{1\dot{2}}k_{2\dot{1}}=k^{\mu }k_{\mu }$, vanishes. Therefore equations
(\ref{conditionsF}), (\ref{definitionsF}) express one constraint only, which
can be written, for example, as: 
\begin{equation}
\left( k_{1\dot{1}}\hat{\alpha}_{\dot{1}}\pi ^{1\dot{1}}+k_{2\dot{1}}\hat{%
\alpha}_{\dot{2}}\pi ^{2\dot{2}}\right) \chi =\left( -k_{1\dot{1}}\hat{\alpha%
}_{\dot{2}}\pi ^{1\dot{2}}-k_{2\dot{1}}\hat{\alpha}_{\dot{1}}\pi ^{2\dot{1}%
}\right) \chi .  \label{constraint}
\end{equation}%
Now we note that Eq. (\ref{constraint}) can be solved, for longitudinal
potentials, by separation of variables, if we put $\chi \left( x\right)
=f\left( x^{0},x^{3}\right) g\left( x^{1},x^{2}\right) $ since $\pi ^{1\dot{1%
}}$, $\pi ^{2\dot{2}}$ act only on $x^{0},x^{3}$ while $\pi ^{1\dot{2}}$, $%
\pi ^{2\dot{1}}$ act only on $x^{1},x^{2}$.

Let longitudinal fields in (\ref{decoupled1}), (\ref{decoupled2}) are
switched off and then crossed fields, obeying $\left[ \pi ^{0},\pi
^{3}]=[\pi ^{1},\pi ^{2}\right] =0$, are turned on. It follows that
equations (\ref{decoupled1}), (\ref{decoupled2}), in presence of such
crossed fields, are the $3\times3 $ equations (\ref{const-s0-1}), (\ref%
{const-s0-2}) obtained from the spin-$0$ DKP equations (\ref{KDP-s0-2}).
Therefore, we can pass directly from equations (\ref{decoupled1}), (\ref%
{decoupled2}), now in crossed fields, to the DKP equations (\ref{KDP-s0-2}).

\section{Discussion}

\label{discussion}

In Section \ref{subequations} we have reviewed the procedure of splitting
the Dirac equation into two $3\times 3$ equations in the non-interacting 
\cite{Okninski2011} as well as interacting case \cite{Okninski2012}. These
equations can be written in covariant form as Dirac equations with some
projection operators but do not contain all components of spinors used in
the splitting. This problem was solved in the free case in Ref. \cite%
{Okninski2014} where we assumed that $\eta _{\dot{B}}\left( x\right) =\hat{%
\alpha}_{ \dot{B}}\,\chi \left( x\right) $, where $\hat{\alpha}_{\dot{B}}$
was a constant spinor. This ansatz for $\eta _{\dot{B}}\left( x\right) $ is
not fully satisfactory since constant spinors do not appear in nature.

To solve these problems we have assumed in the present paper ansatzes (\ref%
{neutrino1F}), (\ref{neutrino2F}), involving neutrino field. The free Dirac
equation has been converted to the set of $3\times 3$ equations (\ref%
{decoupled1F}), (\ref{decoupled2F}), all spinors in the equations appearing
in complete form. There are constraints, (\ref{conditionsF}), (\ref%
{definitionsF}), but they can be easily fulfilled. It follows that equations
(\ref{decoupled1F}), (\ref{decoupled2F}) are equivalent to the spin-$0$ DKP
equations. The inverse transformation, from boson to fermion, in presence of
neutrino field, seems to correspond to the first stage of main channel of
pion decay (\ref{decayF2}) where formation of the intermediate complex state 
$\left( \mu ^{-}\,\bar{\nu}_{\mu }\right)$ has been assumed. Therefore,
kinematics of the reaction products is missing in the present theory. There
is of course another problem, since muon neutrino is massless in this
formalism, although neutrino oscillations indicate that neutrinos are
massive \cite{Beringer2012}.

We have also constructed analogous transition for the Dirac equation in
longitudinal fields. The resulting equations (\ref{decoupled1}), (\ref%
{decoupled2}) are not equivalent to the DKP equation, not in the case of
longitudinal fields. However, switching over to crossed fields in these
equations, after removing neutrino components $\alpha _{\dot{1}}$, $\alpha _{%
\dot{2}}$, we get immediately the DKP equation (\ref{KDP-s0-2}).This
continuous and invertible fermion-boson transition induced by switch-over of
external fields bears some analogy to BCS-BEC transition in ultracold Fermi
gases \cite{Zwerger2012} mentioned in the Introduction. Further analysis of
this problem is needed.

There are several necessary ingredients of these mechanisms: $3\times 3$
equations, Weyl spinor $\alpha _{\dot{A}}\left( x\right)$ and, in the
interacting case, switch-over of external fields. The present mechanism is
an improvement over that of paper \cite{Okninski2014} since it applies to
the interacting case and uses physically meaningful neutrino spinor $a_{\dot{%
B}}\left( x\right) $\ rather than a troublesome constant spinor $\hat{a}_{%
\dot{B}}$. The $3\times 3$ equations may provide a clue to the nature of the
transition mechanism.


\begin{thebibliography}{99}
\bibitem{Garbaczewski1985} P. Garbaczewski, \textit{Classical and quantum
field theory of exactly soluble nonlinear systems}, World Scientific 1985.

\bibitem{Wilczek1990} F. Wilczek, \textit{Fractional statistics and anyon
superconductivity}, World Scientific 1990.

\bibitem{Stone1994} M. Stone, \textit{Bosonization}, World Scientific 1994.

\bibitem{Simulik2005} V.M. Simulik, \textit{What is the Electron?}, Apeiron
2005.

\bibitem{Garbaczewski1975} P. Garbaczewski, Representations of the CAR
Generated by Representations of the CCR. Fock Case, Comm.Math. Phys. 43, 131
(1975).

\bibitem{Jackiw1991} R. Jackiw, V.P. Nair, Relativistic wave equation for
anyons, Phys. Rev. D43, 1933 (1991).

\bibitem{Plyushchay1991} M.S. Plyushchay, Relativistic particle with
torsion, Majorana equation and fractional spin, Phys. Lett. B273, 250 (1991).

\bibitem{Horvathy2010} P.A. Horv\'{a}thy, M.S. Plyushchay, M. Valenzuela,
Supersymmetry between Jackiw-Nair and Dirac-Majorana anyons, Phys. Rev. D81,
127701 (2010).

\bibitem{Okninski2014} A. Okni\'{n}ski, On the Mechanism of Fermion-Boson
Transformation, Int. J. Theor. Phys. 53, 2662 (2014).

\bibitem{Polyakov1988} A.M. Polyakov, Fermi-Bose transmutations induced by
gauge fields, Mod. Phys. Lett. A3, 325 (1988).

\bibitem{Zwerger2012} W. Zwerger (ed.), \textit{The BCS--BEC Crossover and
the Unitary Fermi Gas}, Lecture Notes in Physics 836, Springer-Verlag 2012.

\bibitem{Okninski2011} A. Okni\'{n}ski, Supersymmetric content of the Dirac
and Duffin-Kemmer-Petiau equations, Int. J. Theor. Phys. 50, 729 (2011).

\bibitem{Okninski2012} A. Okni\'{n}ski, Duffin-Kemmer-Petiau and Dirac
Equations -- A Supersymmetric Connection, Symmetry 4, 427 (2012).

\bibitem{Berestetskii1971} V.B. Berestetskii, E.M. Lifshits, L.P.
Pitaevskii, \textit{Relativistic Quantum Theory}, Pergamon 1974.

\bibitem{Bagrov1990} V.G. Bagrov, D.M. Gitman, \textit{Exact solutions of
relativistic wave equations}, Springer 1990.

\bibitem{Beringer2012} J. Beringer et al. (Particle Data Group), Review of
particle physics, Phys. Rev. D 86, 010001 (2012).
\end{thebibliography}
\end{document}